\documentclass[10pt]{iopart}
\usepackage{subfigure}
\usepackage{graphicx}
\usepackage{color}
\usepackage[absolute,overlay]{textpos}
\usepackage{cancel}
\usepackage[mathscr]{euscript}
\usepackage{textcomp}
\usepackage{lmodern}
\usepackage[T1]{fontenc}
\usepackage{breqn}
\usepackage{amssymb}
\graphicspath{ {./Figures/} }

\newcommand{\bv}[1]{\ensuremath{\mathbf{#1}}}

\newcommand{\af}[0]{$\underline{\phi}$ }
\newcommand{\afm}[0]{$\underline{\phi}^i_{max}$ }
\newcommand{\bpf}[0]{$\underline{b}_{\theta}$ }

\begin{document}

\title{Local measurement of error field using naturally rotating tearing mode dynamics in EXTRAP T2R}
\author{R.M. Sweeney\textsuperscript{1}, L. Frassinetti\textsuperscript{2}, P. Brunsell\textsuperscript{2}, R. Fridstr\"{o}m\textsuperscript{2}, 
F.A. Volpe\textsuperscript{1}}

\address{\(^1\) Department of Applied Physics and Applied Mathematics, Columbia University, New York, New York 10027, USA}
\address{\(^2\) Department of Fusion Plasma Physics, School of Electrical Engineering, KTH Royal Institute of
Technology, SE-10044 Stockholm, Sweden}

\begin{abstract}

An error field (EF) detection technique using the amplitude modulation of a naturally rotating tearing mode (TM) is developed and validated in the EXTRAP T2R reversed field pinch. The technique was used to identify intrinsic EFs of $m/n = 1/-12$, where $m$ and $n$ are the poloidal and toroidal mode numbers. The effect of the EF and of a resonant magnetic perturbation (RMP) on the TM, in particular on amplitude modulation, is modeled with a first-order solution of the Modified Rutherford Equation. In the experiment, the TM amplitude is measured as a function of the toroidal angle as the TM rotates rapidly in the presence of an unknown EF and a known, deliberately applied RMP. The RMP amplitude is fixed while the toroidal phase is varied from one discharge to the other, completing a full toroidal scan. Using three such scans with different RMP amplitudes, the EF amplitude and phase are inferred from the phases at which the TM amplitude maximizes. The estimated EF amplitude is consistent with other estimates (e.g. based on the best EF-cancelling RMP, resulting in the fastest TM rotation). A passive variant of this technique is also presented, where no RMPs are applied, and the EF phase is deduced. 

\end{abstract}

\maketitle

\ioptwocol

\section{Introduction}

Error fields are non-axisymmetric fields that result from mis-shaped or misaligned coils and their current feeds. Error fields (EFs) with low-order toroidal harmonics ($n$) and of small relative amplitudes, of the order of $10^{-4}$ of the equilibrium field, are found to cause deleterious effects to plasmas in tokamaks \cite{Buttery1999,Reimerdes,Hender2007}. In reversed field pinches, these field errors are observed to modify the plasma rotation \cite{Frassinetti2012} and can cause rotating tearing modes to lock \cite{Fridstrom2015}.

Various techniques have been demonstrated to measure components of EFs, each with a different set of strengths and weaknesses. Perhaps the most routinely used in tokamaks to date is the low density locked mode onset technique commonly referred to as the "compass scan" technique \cite{Buttery1999,Hender1992}. This method is effective at measuring EFs in low density Ohmic discharges, and has improved stability in high beta H-modes in tokamaks. Vacuum measurements of non-axisymmetric fields can also be made using an in vessel apparatus to accurately measure fields at the location of the plasma, usually during a vessel vent as has been done in DIII-D \cite{LaHaye1991, Luxon2003}. This method is effective at measuring vacuum EFs, but it does not account for deviations of the equilibrium coils due to thermal expansion and $\bv{j} \times \bv{B}$ forces resulting from the plasma current.

The experiments in this work were conducted on the EXTRAP T2R reversed field pinch (RFP) \cite{Brunsell2001}.  Consistent with the RFP literature, perturbations are Fourier analyzed using the form $e^{i(m\theta + n \phi)}$, which leads to the resonance condition when the safety factor takes on a value $q(r)\equiv - m/n$. In these discharges, $1\gg q(0)>0$ on axis and decreases monotonically passing through 0 as the toroidal field $B_T$ reverses at $r/a\approx0.85$, where $a$ is the plasma minor radius, and attains negative values in the range $0 > q(a) \gg -1$ at the plasma edge. As a result of $|q| \ll 1$ everywhere, the main resonant harmonics have $m=1$ and large $|n|$. Taking $m$ always positive, the toroidal harmonics are therefore such that $n<0$ and $n>0$ inside and outside of the field reversal surface respectively. For typical EXTRAP T2R plasma configurations, $m=1$ and $n\leq -12$ are TMs inside the reversal surface, $-12 < n <n_{edge}$ are RWMs, and $n> n_{edge}$ are TMs outside the reversal surface \cite{Drake2005}. The value of $n$ at the edge is given by $n_{edge}=-1/q_{edge}$ (note that since $q_{edge}<0$, $n_{edge}>0$). The most internal resonance is the $m/n=1/-12$ which hosts the TM that will be studied in this work. 

The magnetohydrodynamic (MHD) modes that are often caused by field errors may also be used to diagnose them. Some examples have been implemented in EXTRAP T2R for a stable ideal kink \cite{Volpe2013} and in DIII-D for a stable EF penetration and a saturated locked mode \cite{Shiraki2014,Shiraki2015,Volpe2009}. In EXTRAP T2R, a stable kink was entrained using a resonant magnetic perturbation (RMP) rotating at 50 Hz and the non-uniform rotation of the kink was used to diagnose the resonant component of the intrinsic EF \cite{Volpe2013}. In the DIII-D tokamak, both EF penetration locked modes and saturated locked modes were rotated with RMPs and the EF characterized by their rotation dynamics  \cite{Shiraki2014,Shiraki2015,Volpe2009}. 

Like the stable kink, EF penetration locked mode, and saturated locked mode used for EF correction (EFC), naturally rotating saturated tearing modes (TMs) in EXTRAP T2R are also observed to interact with static resonant fields. The velocity and amplitude of the naturally rotating $m/n=1/-12$ TM in EXTRAP T2R is observed to modulate in the presence of a resonant field that is $\sim10^{-3}$ smaller than the equilibrium field \cite{Frassinetti2010}. 

Here we investigate the use of the modulations in the $m/n=1/-12$ Fourier coefficient of the poloidal field $b_{\theta}^{1,-12}$ of these naturally rotating TMs to measure externally applied resonant fields at the rational surface. The use of saturated TMs allows probing of field components ($m$, $n$) that are not accessible to the driven external kink technique mentioned above \cite{Volpe2013}. This technique is suitable for real-time EFC in plasmas where rotating TMs are present, either because they are unsuppressed or because they are "intentional", as is the case in the tokamak "hybrid" scenario \cite{Sips2002, Luce2003}. When TMs are undesirable (as is often the case), this method is intended to validate, or possibly better the error field correction only when a TM appears. The ability to directly measure resonant fields in H-mode would remove the need to extrapolate measurements made in L-mode.

Although the perturbed TM velocity data appear less sensitive to the external fields than the perturbed amplitude data in these experiments, it should be noted that these are an additional source of external field information. Future EF identification algorithms similar to this one might use perturbed velocity data in place of poloidal field data as cylindrical modeling suggests that this approach might be more applicable to ITER.

This work is organized as follows. In section \ref{sec:setup}, the experimental setup is described. In section \ref{sec:model}, an analytic first-order model describing the time varying width of a naturally rotating TM in the presence of an external resonant field is derived. In section \ref{sec:methods}, the methods used to analyze the experimental data are described. In section \ref{sec:RMP}, the phase identification technique is validated by measuring the known phase of an applied RMP. In section \ref{sec:EF}, the technique is used to measure the amplitude and phase of the unknown $m/n=1/-12$ intrinsic EF of EXTRAP T2R. In section \ref{sec:discussion}, limitations of the model and the analysis are discussed, and separately the potential of this and similar techniques for use in ITER are briefly investigated. Appendix A is dedicated to relating the two coordinate systems used in this work, and Appendix B to consolidating all angle parameters in a single table with brief definitions. The reader is encouraged to refer to Appendix B as needed.

\section{Experimental setup}
\label{sec:setup}

EXTRAP T2R is a reversed field pinch with major and minor radii $R=1.24$ m and $r=0.183$ m. Magnetic feedback on unstable modes in EXTRAP T2R allows plasma discharges of duration $\approx70-90$ ms. In the experiments presented here we have $I_p \approx 90$ kA, $n_e\approx(0.5-1.0) \times 10^{19}$ m$^{-3}$, and $T_e \approx 200-400$ eV. The stainless steel vacuum vessel is located at $r_v=0.192$ m with a resistive diffusion time constant for the $m/n=1/-12$ mode of  $\tau_v \approx 58$ $\mu$s as reported in \cite{Polsinelli2015}. Outside of the vacuum vessel at $r_w=0.198$ m is a concentric copper shell consisting of two layers with a total thickness of $\delta_w=1$ mm and a resistive diffusion time constant of $\tau_w \approx 2 \times 6.3 \approx 13$ ms \cite{Brunsell2001}, which is in good agreement with the theoretical value $\tau_w = \mu_0 \sigma_w r_w \delta_w \approx 13.8$ ms. 

Tearing mode dynamics are measured by 4 (poloidal) $\times$ 64 (toroidal) magnetic probes \cite{Frassinetti2007,Frassinetti2010} measuring the local poloidal field $b_{\theta}(\bv{x},t)$ at position $\bv{x}$ and time $t$. The probes are positioned between the vacuum vessel and the copper shell. The rotation frequencies of interest are such that $\omega^{1,-12} \sim \tau_v^{-1}$ and therefore compensation of vessel eddy currents is important. To decouple the effect of eddy currents, the vessel is assumed to be a first-order low-pass filter with transfer function $H(f) = 1 + i f/f_c$ for signals with frequency $f$ and where $f_c=2.7$ kHz \cite{Polsinelli2015} is the cutoff frequency (see reference \cite{Frassinetti2007} for details). The magnetics are compensated by applying an inverse filter.  An additional array of 4  $\times$ 32  saddle loops located outside of the copper shells are used to measure radial magnetic fields on slow timescales (less than or equal to $\tau_w^{-1}$), which are suppressed by feedback in normal operation by a complementary array of 4  $\times$ 32  actuator coils. The revised intelligent shell algorithm \cite{Frassinetti2011, Olofsson2009} takes the radial field from the saddle loops as input and can be programmed to suppress all harmonics, or to fix chosen harmonics to a given set-point, as will be done here with the $m/n=1/-12$ harmonic. Choosing a set amplitude and phase in this algorithm for the $m/n=1/-12$ field is different from applying a constant $m/n=1/-12$ field with the actuator coils as the feedback takes into account the plasma response, driving the total field (RMP and plasma response) toward the requested set-point.

\subsection{Notations for rotation frequencies and toroidal phases of TMs}
\label{sec:notations}

When working with TMs with $|n| \neq 1$, two frequencies and corresponding phases may be used to describe their toroidal motion. A magnetic sensor fixed to the vessel will measure a field that oscillates like $\sin(\delta - n \omega^{1,-12} t)$ where $\omega^{1,-12}$ is the toroidal plasma rotation frequency at the rational surface, assuming the TM is entrained in the plasma flow, and $\delta$ is an arbitrary offset depending on the sensor position and the initial position of the TM. A second frequency can be defined which treats one sinusoidal oscillation in the magnetics as one period, and is thus defined as $\omega \equiv n \omega^{1,-12}$. Correspondingly, we can time integrate this relationship to find $\phi \equiv n \phi^{1,-12}$. All references to rotation frequencies $\omega$ and toroidal angles $\phi$ in following sections will refer to $\omega$ and $\phi$ as defined here (i.e. $\omega^{1,-12}$ and $\phi^{1,-12}$ will not be used outside of section \ref{sec:setup}). Note that this applies to \textit{all} angles and frequencies, including the TM, the applied RMP, and the intrinsic EF. 

The $\phi$ position of a continuous field is ambiguous, so here we will explicitly define what the $\phi$ position of all $m/n=1/-12$ fields in this work means. Assuming the existence of some $m/n=1/-12$ field that we call $A$, we take the toroidal angle where the radial field is maximally directed outward at the outboard mid-plane to be $\phi_A$. Note that although this point is twelve times degenerate in the $\phi^{1,-12}$ coordinate system on the domain $0 \le \phi^{1,-12} < 2\pi$, it is unique in the $\phi$ coordinate system on the domain $0 \le \phi < 2\pi$ (recall that $\phi \equiv n \phi^{1,-12}$). 

Finally, all field magnitudes in what follows will refer to the $m/n=1/-12$ Fourier coefficient. Superscripts 1,-12 will appear in few places for clarity, but the reader should interpret all field magnitudes without superscripts as the $m/n=1/-12$ Fourier coefficient. 

\subsection{Discharge design}

By $t=20$ ms, the transients associated with startup have decayed and the plasma has settled into an approximately constant equilibrium. At this time, we program the feedback to apply a static $m/n=1/-12$ RMP of constant amplitude $B_{RMP}$. The RMP is applied between $t=20-40$ ms, and is the time window over which all of the following analysis is done.

The phase of the applied RMP $\phi_{RMP}$ is changed between shots providing 20 ms of interaction between the TM and RMP for each $\phi_{RMP}$.  A scan of 9 phases spanning 0 to $2\pi$ was completed with an amplitude of $B_{RMP}=2$ G, and a scan of 7 phases spanning 0 to $2 \pi$ was completed for a 50\% stronger RMP with $B_{RMP}=3$ G. The results of these two scans are detailed in section \ref{sec:RMP} to demonstrate phase identification of a known RMP. To motivate the technique that is used, we now introduce a simple analytic model to describe the expected TM behavior.

\section{Model of fast TM/EF interaction}
\label{sec:model}

While in reference \cite{Frassinetti2010} the Modified Rutherford Equation (MRE) and equation of motion are solved numerically, here we will model only the island width behavior using the MRE and seek an analytic first-order perturbation expansion solution. Although we will see that the equation of motion and the MRE are coupled, the effect of the coupling is expected to be second order and is omitted from this model. 

The estimated TM width $W_0$ is found to be comparable with the linear layer width. This implies that the TM is weakly nonlinear as parameterized by $\lambda \sim (\delta_s/ W_0)^{3/2} \sim 0.3$ (see equation 101 in reference\cite{Fitzpatrick1993}). Despite being weakly nonlinear, the Modified Rutherford Equation is used to describe the observed TM amplitude dynamics. 

\subsection{Modified Rutherford Equation}

Similar to \cite{Frassinetti2010}, we take the following form for the Modified Rutherford Equation (MRE) not including the bootstrap current term (that is, describing classical TMs, not neoclassical TMs),

\begin{equation}
	\frac{\tau_R}{r_s} \frac{d W}{dt} = \Delta'(W)r_s + \Gamma \frac{W_v^2}{W^2}  e^{i \Delta \phi(t)}
	\label{eq:MRE}
\end{equation}

where $\tau_R = \mu_0 r_s^2/\eta$ is the resistive diffusion timescale, $r_s$ is the minor radius of the rational surface, $W$ is the island width, $\Delta'(W)$ is the classical stability index which depends on the island width, $\Gamma$ is a function that depends on the geometry of the fields and the boundary conditions at conducting surfaces (see Appendix A in reference \cite{Frassinetti2010}), $W_v$ is the vacuum island width driven by an external resonant field, and $\Delta \phi(t)$ is the toroidal angle between O-points of the vacuum island and plasma island (the vacuum island is found by superimposing the external field on the equilibrium field in vacuum). 

The neoclassical bootstrap term often included in modeling of high beta devices is considered negligible here as the pressure is relatively low and the inverse aspect ratio is small. Complex notation is used in equation \ref{eq:MRE} though only the real part has physical meaning, and throughout this work we will only consider the real part of all equations and all quantities, including the island width $W$. 

As in \cite{LahayeNF15, Fitzpatrick1993}, we express $\Delta'(W)$ as a constant plus a linear term in the island width,

\begin{equation}
	\Delta'(W) r_s = C_0 - C_1 \frac{W}{r_s}
	\label{eq:delta'}
\end{equation}

where $C_0$ and $C_1$ are dimensionless constants. We will be concerned with saturated rotating islands and relatively weak external fields such that $W > W_v$. We will therefore take $\Gamma (W_v/W)^2 \sim \epsilon$, where $\epsilon$ is a small quantity used for ordering. The zeroth order saturated island width is given by, 

\begin{equation}
	W_0 = \frac{C_0}{C_1} r_s
	\label{eq:zeroWidth}
\end{equation}

We are now interested in small perturbations about $W_0$. Substituting $W = W_0 + \epsilon W_1$ in equation \ref{eq:MRE}, and retaining only terms of order $\epsilon$, we find,

\begin{equation}
	\frac{\tau_R}{r} \frac{d W_1}{dt} = - C_1 \frac{W_1}{r} + \Gamma \frac{W_v^2}{W_0^2}  e^{i \Delta \phi(t)}
	\label{eq:perturbedMRE}
\end{equation}

where we have used the assumption that $\Gamma (W_v/W)^2 \sim \epsilon$, and kept only the zeroth order Taylor expansion of this term.

The external field term involving $\Delta \phi(t)$ is responsible for coupling the MRE and the equation of motion in this high-frequency regime where $\omega_o \gg \tau_w^{-1}$ (at lower frequencies, couplings  also occur due to resistive eddy currents in the wall \cite{Finn1994}). Although the rotation of the TM is not uniform in the presence of an external field, the perturbations to $\phi(t)$ are observed to be relatively small. As we have already taken the second term on the right-hand-side of equation \ref{eq:perturbedMRE} to be order $\epsilon$, the effect of this small perturbation in $\Delta \phi(t)$ is order $\epsilon^2$ and therefore omitted. That is to say that in this model for the perturbed island width, we assume uniform rotation. Taking $\Delta \phi(t) \approx \omega_o t$, where the arbitrary initial phase is chosen such that $\Delta \phi(0) = 0$ (the TM and EF are aligned at $t=0$), we find,

\begin{equation}
	\frac{W_1(t)}{r} =  \left( \frac{W_v}{W_0} \right )^2 \frac{\Gamma}{\tau_R} \left [ \frac{ (C_1/\tau_R)  - i \omega_o  }{ \omega_o^2 + (C_1/\tau_R)^2} \right ]  e^{i \omega_o t}
	\label{eq:oneWidth}
\end{equation}

From equation \ref{eq:oneWidth} we see that the oscillation in island width maximizes on the domain $\Delta \phi=[0, \pi/2]$ (recall that $\Delta \phi(t)$ and $\omega_o t$ are approximately interchangeable). The exact phase, referred to as $\Delta \phi_{max}$, depends on the relative sizes of $C_1/\tau_R$ and $\omega_o$.

We now write the full time dependent island evolution as, 

\begin{equation}
	W(t) = W_0 + W_1(t)
	\label{eq:w_t}
\end{equation}

We know that $W \propto \sqrt{b_{\theta}}$ where $b_{\theta}$ here is the $m/n=1/-12$ Fourier coefficient \cite{Fitzpatrick1993}. Thus, taking the square of equation \ref{eq:w_t} and omitting terms of order $W_1^2$, 

\begin{equation}
	b_{\theta}(t) = b_{\theta 0} +  b_{\theta 1} 
	\label{eq:bp}
\end{equation}

where $b_{\theta 0} \propto W_0^2$, and $b_{\theta 1}(t) \propto 2 W_0 W_1(t)$. We now have a model for the time dependent poloidal field of the TM. 

In summary, in this section we have seen that $b_{\theta}$ oscillates once per rotation period, and the phase at which $b_{\theta}$ is maximum contains information on the toroidal phase of the EF in the lab frame.

\section{Methods}
\label{sec:methods}

\subsection{Data filtering}

First, the raw $dB/dt$ data are time integrated and compensated for wall eddy currents as discussed in section \ref{sec:setup}. From the model developed in section \ref{sec:model}, we expect the amplitude of the $m/n=1/-12$ TM to oscillate with frequency $\omega_o$. The amplitude and phase of the TM are also affected by other mechanisms, in particular the sawtooth oscillations. The TM dynamics associated with the sawtooth oscillations are not of interest here, and therefore the data are filtered to remove these transient events. A high-pass filter is implemented by simply subtracting from the time-integrated magnetics, a copy of the signals smoothed over the characteristic rotation period time. A second high-pass filter is then implemented by Fourier transforming into the frequency domain, zeroing the spectrum below 30 kHz, and performing an inverse transform on the truncated spectrum. 

At this point the data are Fourier analyzed in space, and all following analysis will refer to the $m/n=1/-12$ TM. 

The integrated $m/n=1/-12$ field amplitude is now low-pass filtered with a cutoff frequency of 110 kHz to remove high-frequency noise. 

Next, a \textit{rotation filter} is applied, which is based upon the observed TM rotation, as measured by the magnetics. The data must satisfy the following two conditions to be included in the analysis; (1) $\phi$ must complete full rotations from 0 to $2\pi$, and (2) the time derivative of $\phi$ must remain above a specified threshold. Condition (1) removes times when the amplitude modulation becomes larger than the unperturbed amplitude, where the mode repeatedly appears and disappears \cite{Frassinetti2010}. Condition (2) ensures that the mode is not locked. Further, conditions (1) and (2) must be satisfied for a duration $\tau \geq 0.1$ ms, which ensures that the majority of the rotation data are temporally isolated from nearby locking events. After the high-pass and rotation filters, we notate the resulting data \bpf and \af. Note that we have dropped the superscripts denoting 1,-12 on \bpf. 

\subsection{Feature extraction}

 We now proceed with the physical analysis of these signals. A semi-empirical algorithm is developed to process the \bpf and \af  data, returning an estimate of the external resonant field phase (e.g. the error field phase $\phi_{EF}$, the RMP phase $\phi_{RMP}$, or their superposition $\phi_{EF+RMP}$) with which the TM interacts. The algorithm is based on the premise that \bpf is expected to reach a single maxima during each period of rotation (see equation \ref{eq:bp}).  

\begin{figure}[h]
	\centering
	\includegraphics[scale=0.45]{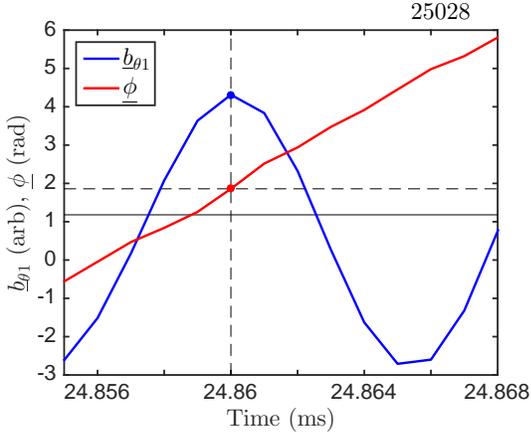}
	\caption{
	Shown are the time-varying amplitude of the poloidal field (blue) and the phase of the TM (red) in the presence of a $B_r=2$ G RMP during
	rotation period $i$ in shot 25028. The vertical dashed line intersects the point where the poloidal field is maximized $b_{\theta, max}^i$ (blue point) and 
	the TM phase at which this maxima occurs \afm (red point). The solid horizontal line shows the phase of the applied RMP. 
	}
	\label{fig:phiMax}
\end{figure}

Motivated by this discussion, we look for the phase of the TM at which \bpf is maximized in each rotation period. Figure \ref{fig:phiMax} shows $\underline{b}_{\theta 1}$ (i.e. only the time-varying portion of the TM Fourier coefficient) and \af during one rotation period in the presence of a $B_r=2$ G RMP applied with static phase. The vertical dashed line intersects both the maximum of $\underline{b}_{\theta 1}$, and the phase \af at which this maximum is achieved. We refer to this phase in rotation period $i$ as $\underline{\phi}^i_{max}$. The process of identifying \afm is then repeated for all rotation periods, providing $\sim 1,500$ values of \afm for a given phase of the  external resonant field (a typical shot contains 1 to 2 thousand rotation periods).

\begin{figure}[h]
	\includegraphics[scale=0.75]{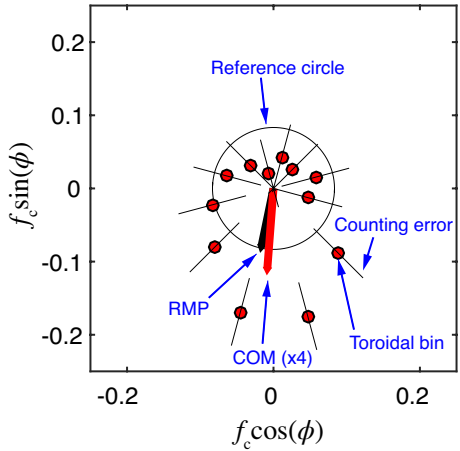}
	\caption{
	All \afm data for a given shot are displayed on a polar histogram. This polar plot spans $30^\circ$ in real-space, such that an $n=12$ perturbation appears like
	an $n=1$. The radius of each red point corresponds to the fraction of the total counts $f_c$
	in the given toroidal angle bin. The line segment intersecting each red point quantifies the Poisson counting 
	error. The black arrow is a phasor representation of the applied RMP. The red arrow
	points to the center-of-mass (COM) of the red points (scaled by 4 for visual purposes). If all \afm data were distributed evenly among the bins,
	then the red points would appear on the black reference circle.
	}
	\label{fig:polarExample}
\end{figure}

All estimates of \afm for a given external field phase are now histogrammed in toroidal angle. These histograms are mapped onto a polar plot, as shown in figure \ref{fig:polarExample}. Note that this polar plot spans 30 degrees in real space, such that an $n=12$ perturbation will appear $n=1$ on this plot. Each bin in figure \ref{fig:polarExample} is represented by a red point. The counts in each bin are normalized by the total number of counts across all bins, thus representing the fraction of counts $f_c$.  The value of $f_c^j$ for bin $j$ determines the radial distance of the red point from the origin.  A TM that behaved exactly according to our model (equation \ref{eq:bp}) would produce a distribution where all \afm fall within a single bin $j$ with $f_c^j=1$. The distribution in figure \ref{fig:polarExample} is clearly much broader than this expected distribution, suggesting that our model is too simple to capture all of the physics here. Possible explanations of the broadened distribution will be discussed in section \ref{sec:discussion}. 

However, although simplistic, the model in section \ref{sec:model} might explain why the distribution of \afm does peak at a specific $\Delta \phi$ (i.e. $\Delta \phi_{max}$). As our physical model does not describe the shapes of these polar histograms, we employ a simple approach to extract the external field phase from the histogram data. We attribute equal "mass" to each data point in figure \ref{fig:polarExample} and calculate the center of mass. The phasor that points to it (shown in red in figure \ref{fig:polarExample}) will be called the "centroid phasor". The toroidal angle of this centroid phasor is then used as an estimate of the external field phase. The black phasor in figure \ref{fig:polarExample} shows the phase of the applied RMP. The magnitude of the black phasor is arbitrary and therefore should not be compared with the magnitude of the red phasor. 

\subsection{Statistical uncertainty}

A Monte-Carlo technique is used to quantify the uncertainty in the centroid phase: each bin value is varied about the measured value according to a normal distribution. The standard deviation of such distribution is given by the Poisson counting error, shown by the line segments intersecting the red histogram data in figure \ref{fig:polarExample}. The histograms are perturbed in this way 200 times, and the phase of the centroid recalculated for each perturbed distribution. The standard deviation of the 200 centroid phases is then used to provide an uncertainty in the measured centroid phase. This is not shown in figure \ref{fig:polarExample}, but will appear as an error bar in figures \ref{fig:2GCentroid} and \ref{fig:0p5G_RMP}.

\section{Passive EF phase identification}
\label{sec:RMP}

\begin{figure*}[t]
	\centering
        	\includegraphics[scale=0.52]{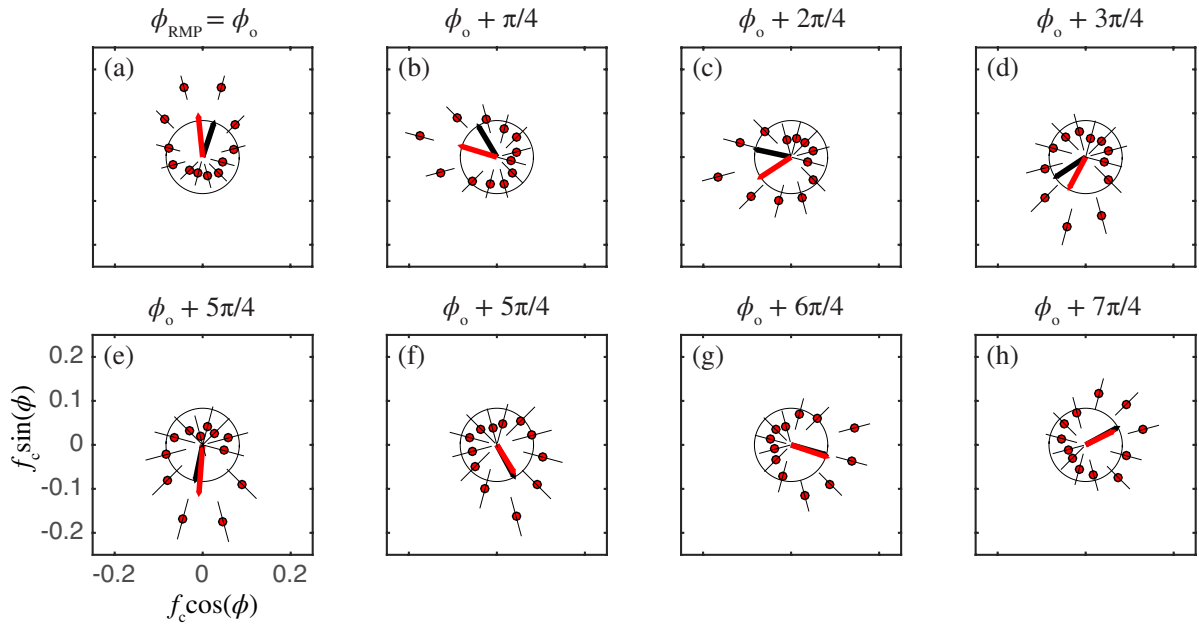}
	\caption{
	Polar histograms plotted for an 9 shot scan (8 shown for visual purposes) of the phase of an applied RMP with amplitude $B_r=2$ G. See figure \ref{fig:polarExample} for a 
	detailed explanation of an enlarged version of figure \ref{fig:polarArray}e. The center-of-mass centroids in these subfigures (red arrow) are scaled by 4 for visual purposes. 
	(a) Polar histogram for $\phi_{RMP}=0.375\pi$ $\equiv \phi_o$ (note that this corresponds to $\Phi_{RMP} = 0$ ; see Appendix A and section \ref{sec:setup}). 
	(b-h) Polar histograms for each $\pi/4$ increment of the toroidal position of the RMP completing a $2 \pi$ scan. Note that the 
	polar plots here span $30^\circ$ in real space (i.e. what appears $n=1$ here is $n=12$ in real space). 
	}
	\label{fig:polarArray}
\end{figure*}

Here the object of the measurement - the external field - is the resultant of the residual $m/n=1/-12$ EF and of applied RMPs of the same $m$ and $n$. The measurement is passive in the sense that it is not necessary to actively probe the system with applied RMPs. For the sake of validation, we will apply known RMP fields and measure them with this technique. In this section we report the results from two sets of discharges where the toroidal phase of a $m/n=1/-12$ RMP is varied between discharges, and the amplitude of the RMP is changed between the two sets. In each discharge, the RMP is applied between $t=20$ and 40 ms with static phase. All other harmonics of the slowly varying radial field are suppressed by the feedback.

The results of the $B_r=2$ G RMP phase scan are shown in figure \ref{fig:polarArray}. The applied RMP starts at $\phi_{RMP}=\phi_o \approx 0.33 \pi$  and is incremented in steps of $\pi/4$ until completing a full scan (note $\phi_o$ corresponds to $\Phi=0$ in the active coil coordinate system used by the feedback; see appendix A). Recall that a complete scan of $\phi$ from 0 to $2\pi$ here corresponds to a span of $2 \pi/12$ in real space. The red centroid phasors, shown by the red arrows in the subpanels of figure \ref{fig:polarArray}, are observed to lead the black RMP phasors by up to $\Delta \phi \approx \pi/4$. Note the systematic increase of $\Delta\phi$ from subpanel (c) to (h), followed by a decrease from (h) to (a), (b), and (c).

\begin{figure}[h]
	\centering
	\includegraphics[scale=0.6]{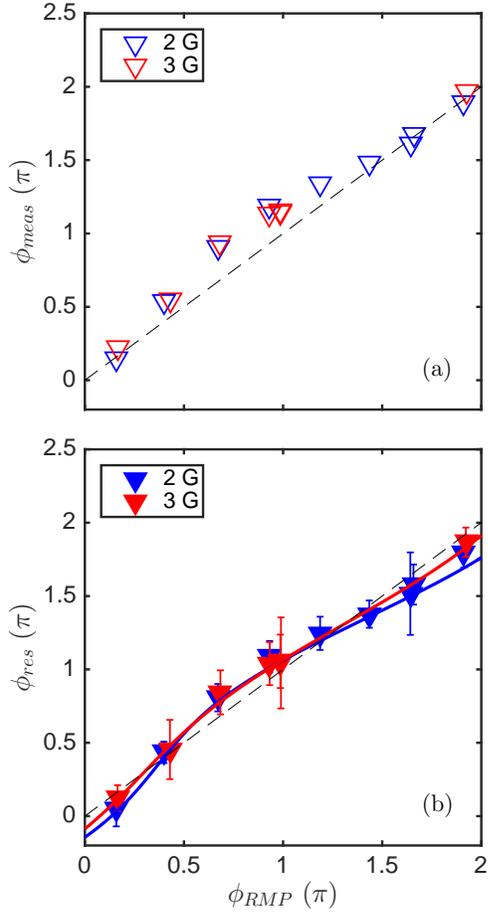}	  
	\caption{
	(a) The measured phase of the centroids $\phi_{meas}$ from figure \ref{fig:polarArray} are plotted as a function of the applied RMP phase. 
	The dashed line shows where $\phi_{meas} = \phi_{RMP}$. 
	(b) The data from (a) are shifted down by $\Delta \phi_{max} = 0.1 \pi$ (referred to as $\phi_{res}$ after this shift), and are observed to oscillate 
	about the dashed line $\phi_{res} = \phi_{RMP}$. Both datasets are fit to equation \ref{eq:phiRes} shown by the solid
	curves. These fits provide estimates of the EF amplitude and phase which are reported in figure \ref{fig:intrinsicEF}.
	}
	\label{fig:2GCentroid}
\end{figure}

In figure \ref{fig:2GCentroid}a, the measured centroid phase is plotted as a function of the RMP phase for the $B_{RMP}=2$ G scan and $B_{RMP}=3$ G scan (open blue and red triangles). It is clear that the measured phase is tracking the phase of the RMP up to a small constant offset (seen by the general vertical shift of the data above the dashed line) and a systematically varying offset (seen by the approximately sinusoidal variation of the data).

\section{Active amplitude and phase identification of intrinsic EF}
\label{sec:EF}

Here, we consider the amplitude and phase of the RMP to be known, and instead search for the existence of an unknown resonant field (e.g. an EF). In addition to identifying the phase of the unknown field as in section \ref{sec:RMP}, the presence of a known RMP will allow us to also measure the amplitude of the unknown EF.  The two sets of shots from section \ref{sec:RMP} ($B_{RMP} =$2, 3 G) will be used again here (now considering the applied RMPs as known), as well as three additional sets with $B_{RMP} = 0$, 0.5, and 1 G. The discharge with no applied RMP measures only the phase of the intrinsic EF for the reasons just discussed (a mathematical justification will be given later in this section).

Recall that all analysis reported in this work refers to the time-interval between $t=20$ and 40 ms in the shot cycle, where we expect the plasma to be in equilibrium. Assuming that the currents in the equilibrium coils and in the plasma are unchanging during this time, we might also expect that the intrinsic EF is static.  

The superposition of the RMP and intrinsic EF phasors produces a resultant phasor, with toroidal phase given by,

\begin{dmath}
	\phi_{res} = \arg \left \{ B_{RMP} e^{i\phi_{RMP}} + B_{EF} e^{i \phi_{EF}} \right \}
	\label{eq:phiRes}
\end{dmath}

where $B_{EF}$ and $\phi_{EF}$ are the magnitude and phase of the unknown intrinsic EF, and $arg\{\}$ is the argument function that returns the angle between the positive real axis and the phasor in the complex plane. In general, the measured phase of the TM maximum, $\phi_{meas}$, will differ from the just defined $\phi_{res}$ by an amount $\Delta \phi_{max}$:

\begin{equation}
	\phi_{meas} = \phi_{res} + \Delta \phi_{max}
	\label{eq:phiMeas}
\end{equation}

This model is fit to the $B_{RMP}=2$ G data in figure \ref{fig:2GCentroid}a, where $B_{EF}$, $\phi_{EF}$, and $\Delta \phi_{max}$ are free fitting parameters. From this fit, we estimate $\Delta \phi_{max} \approx 0.1 \pi$. This value of $\Delta \phi_{max}$ is used to convert all $\phi_{meas}$ measurements to $\phi_{res}$ in the remainder of this work. The data are shifted downward by $\Delta \phi_{max}$ and the model fits are shown in figure \ref{fig:2GCentroid}b. 

Equation \ref{eq:oneWidth} shows that the value of $\Delta \phi_{max}$ has implications on the relative sizes of $C_1/\tau_R$ and $\omega_o$. Although interesting for validation of theory, these implications are not important to the present analysis but will be discussed later in section \ref{sec:discussion}. 

\begin{figure}[h]
	\includegraphics[scale=0.5]{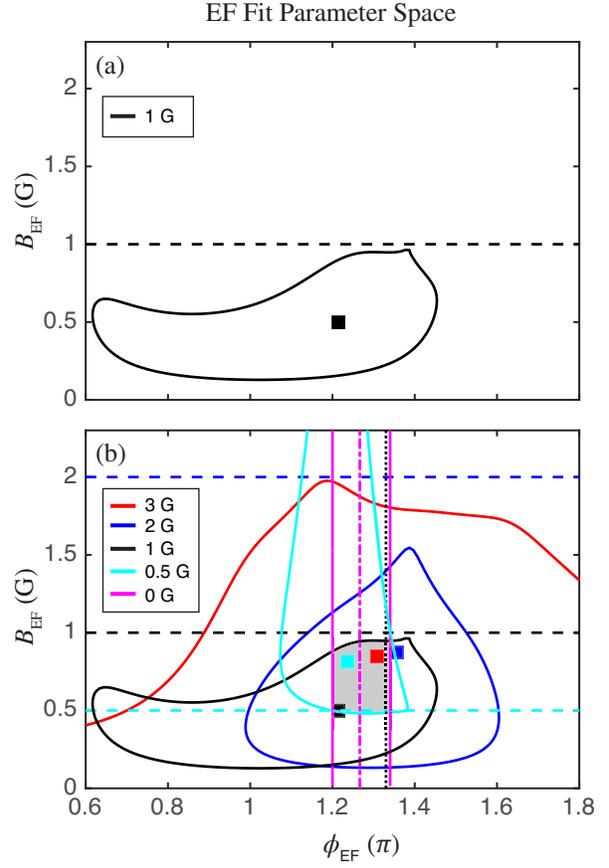}  
	\caption{
	(a) The best fit EF amplitude and phase from the $B_{RMP}=1$ G shot scan are shown by the black square. The black contour bounds the one sigma confidence region within which the 	EF amplitude and phase are expected to exist. The horizontal black dashed line shows the amplitude of the RMP. Note that the black contour is bounded from above by this dashed line.
	(b) Five scans of $\phi_{RMP}$ at a given $B_{RMP}$ (red 3 G, blue 2 G, black 1 G [same as (a)], cyan 0.5 G, and magenta 0 G).  The magenta square is replaced by 
	a magenta vertical dash-dotted line as the measurement with $B_{RMP}=0$ is not sensitive to the EF amplitude (see equation \ref{eq:phiRes}). The gray region
	shows where the error field is predicted in $B_{EF}$ vs. $\phi_{EF}$ space. The vertical dotted black line
	shows the position of the RMPs in figure \ref{fig:EFRotation}. 
	}
	\label{fig:intrinsicEF}
\end{figure}

For redundancy, equation \ref{eq:phiMeas} is fit to five different shot scans (with $B_{RMP} = 0$, 0.5, 1, 2, and 3 G), providing five independent estimates of the intrinsic EF amplitude and phase (note that $\Delta \phi_{max}$ is now known, and no longer a fit parameter). The results of this analysis are shown in figure \ref{fig:intrinsicEF} where squares show each individual estimate. Such estimates are the loci, in the $\phi_{EF}$, $B_{EF}$ plane, where the reduced chi-square $\chi_r^2$ is minimum ($\chi_r^2 = \chi_{r,min}^2$). The contours in figure \ref{fig:intrinsicEF}, on the other hand, correspond to $\chi_r^2=\chi_{r,min}^2 + 1$. Hence, they bound the regions where the fit-parameters are known to within one standard deviation, $\sigma$ (see reference \cite{Bevington} for details).

The set of discharges in which no RMPs are applied (magenta) show the highest sensitivity to $\phi_{EF}$, constraining the phase to $\phi_{EF} = (1.28 \pm 0.08) \pi$ radians. Despite the high phase sensitivity, the amplitude of the EF cannot be deduced when $B_{RMP}=0$, as it can be recognized from equation \ref{eq:phiRes} that $B_{EF}$ scales both components of the complex phasor, and therefore does not change $\phi_{res}$. For this reason, a magenta vertical dotted line marks the phase of $\chi_{r,min}^2$ for the no RMP scan.  The $B_{RMP}=3$ G scan (red) has the lowest sensitivity to $\phi_{EF}$ as within one sigma, it cannot constrain $\phi_{EF}$ at all. 

\begin{figure}[h]
	\centering
	\includegraphics[scale=0.5]{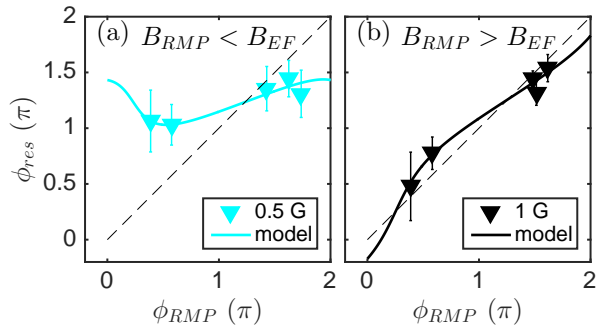}  
	\caption{
	(a) Four shot scan of a 0.5 G RMP, varying $\phi_{RMP}$ shot-to-shot. The blue curve is a fit of equation \ref{eq:phiMeas}, where the minimum
	$\chi_r^2$ occurs for $B_{EF} \approx 0.6$ G and $\phi_{EF}\approx1.3\pi$. Note that the values of $\phi_{meas}$ span less than 
	$\pi$ radians, suggesting that the EF is greater than 0.5 G.  
	(b) Five shot scan of $\phi_{RMP}$ with a constant $B_r=1$ G. The black curve is a fit of equation \ref{eq:phiMeas}, where the minimum
	$\chi_r^2$ occurs for $B_{EF} \approx 0.9$ G and $\phi_{EF}\approx1.25\pi$. The values of $\phi_{meas}$ span the whole polar plane, suggesting
	that this RMP is greater than the EF.  
	The dashed black line in both (a) and (b) shows where $\phi_{meas} = \phi_{RMP}$. 
	}
	\label{fig:0p5G_RMP}
\end{figure}

Using the estimated parameters from fitting equation \ref{eq:phiMeas} to a single toroidal scan provides at best $\pm0.4$ G EF amplitude resolution, as shown by the black contour for the $B_{RMP}=1$ G scan. Nonetheless, an important observation can be made from figure \ref{fig:intrinsicEF} that increases the amplitude resolution: when the applied RMP is smaller than the unknown field, the resulting measured data span a range of less than $\pi$ radians. Therefore, we can conclude from the range of the measured data whether the unknown field is larger than, or smaller than the RMP. For example, the $\phi_{res}$ data from the 0.5 G phase scan span less than $\pi$ radians, as seen in figure \ref{fig:0p5G_RMP}a. This suggests that $B_{EF}> 0.5$ G. Corroborating this, the $1 \cdot \sigma$ contour shown in cyan in figure \ref{fig:intrinsicEF} is lower-bounded by $B_{EF}\approx0.5$ G. The $\phi_{res}$ data from the 1 G phase scan span more than $\pi$ radians as seen in figure \ref{fig:0p5G_RMP}b, suggesting that $B_{RMP} > B_{EF}$. Again, the corresponding black contour in figure \ref{fig:intrinsicEF} corroborates this observation as we see that the dashed black horizontal line bounds the contour from above, meaning that this fit for $B_{RMP}=1$ G is not consistent with $B_{EF} > 1$ G. 

Multiple RMP scans where the RMP amplitude is varied above and below the unknown amplitude $B_{EF}$ constrains the amplitude better than a single scan alone. All five solid contours in figure \ref{fig:intrinsicEF} share a small region of intersection highlighted in gray. Note that since this region is bounded by three scans (black, cyan, and magenta), only these three scans are necessary to constrain the EF. From this region of intersection, we estimate that the EF has phase $\phi_{EF} = (1.28 \pm 0.08)\pi$  and magnitude $B_{EF} = 0.7 \pm 0.2 $ G. 

\subsection{Independent verification of EF estimate}

A scan of RMP amplitude at a constant phase of $\phi_{RMP} = 0.33\pi$  comes close to cancellation of this predicted $m/n=1/-12$ EF and is used as an independent verification of this prediction. This scan consists of $B_{RMP}$ amplitudes of -2, -1, 0, 0.5, 1.5, and 2 G. As just discussed, according to our EF estimate, the EF is believed to be located at $\phi_{EF} \approx 1.3 \pi$  with an amplitude of $B_{EF}\approx 0.7$ G. The field which would cancel this predicted EF would be located at $\phi_{RMP} = 0.3\pi$  and with an amplitude of $B_{RMP} = 0.7$ G. Although this preferred canceling field is not included in the scan just described, the discharge with $B_{RMP}=0.5$ and $\phi_{RMP}=0.33\pi$ is close to the desired field. 

\begin{figure}[h]
	\includegraphics[scale=0.47]{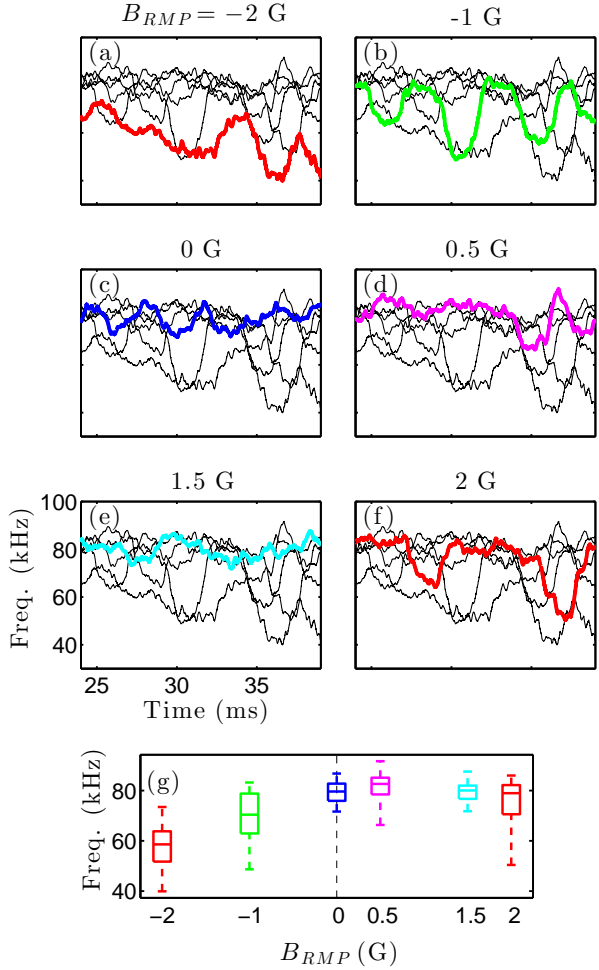}  
	\caption{
	Tearing mode rotation frequency over a 6 shot RMP amplitude scan at $\phi_{RMP} = 0.33 \pi$. 	
	(a-f) Rotation frequency of the $m/n=1/-12$ TM in the presence of the RMP amplitude specified in the title and shown in color, as well as the five other
	frequency traces for comparison in black. All frequency data are smoothed over 2 ms. 
	(g) Box-and-whisker plot of 2 ms smoothed frequency data from $t=24-39$ ms for each amplitude of the applied RMP. Bottom 
	and top of each box represent the $25^{th}$ and $75^{th}$ percentiles, the horizontal line inside the box is the median, and the
	dashed and capped lines extending from the box mark the extrema of the distribution. The vertical dashed black line at $B_{RMP}=0$ G
	is a guide for the eye.
	}
	\label{fig:EFRotation}
\end{figure}

For each of the shots in this scan we observe the frequency of TM rotation during the period when the RMP is applied. Non-axisymmetric fields in EXTRAP T2R are known to apply DC braking torques to the TM through a nonlinear interaction of the field with the amplitude modulated TM \cite{Frassinetti2012}. We therefore expect the TM rotation frequency to be highest when the external resonant field (i.e. the intrinsic EF plus the RMP) is smallest. Each subfigure \ref{fig:EFRotation}a-f shows the TM rotation frequency for a given RMP amplitude in color plotted on top of the four other shots in black for comparison.

Subfigure \ref{fig:EFRotation}g is a "box-and-whisker" plot that summarizes subfigures \ref{fig:EFRotation}a-f. The bottom and top of each box mark the frequency of the $25^{th}$ and $75^{th}$ percentiles and the horizontal line inside the box marks the median. The dashed and capped lines extending out of the top and bottom of each box mark the extreme values of each distribution. 

It is evident that the box-and-whisker plot is not symmetric about $B_{RMP}=0$ G, but rather the vertical line of symmetry appears to occur somewhere between $B_{RMP}=0$ and 1.5 G. For example, the $B_{RMP}=-2$ and -1 G cases are characterized by significantly slower TM rotation compared with the +1.5 and +2 G cases. The fastest rotation is obtained for $B_{RMP}=0.5$ G. These observations are consistent with the prediction that the EF lies somewhere in the intersection region of figure \ref{fig:intrinsicEF}, and thus provides greater confidence in this prediction.

\begin{figure}[h]
	\includegraphics[scale=0.5]{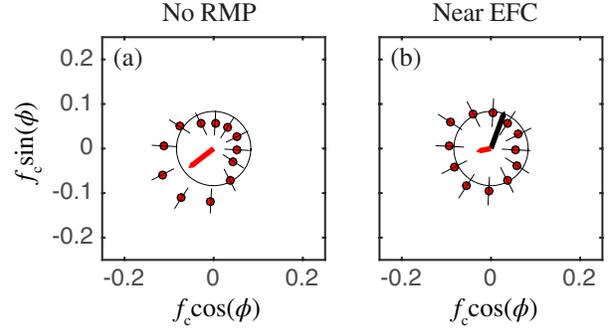}  
	\caption{
	Polar histograms of \afm for two of the shots in figure \ref{fig:EFRotation}. (a) A shot with no RMP. 
	(b) A shot with an RMP (black arrow) that cancels a field in the predicted range of the EF in
	figure \ref{fig:intrinsicEF} (i.e. $B_{RMP}=0.5$ G and $\phi_{RMP} = 0.33\pi$). 
	(a) and (b) here correspond to figures \ref{fig:EFRotation}c and \ref{fig:EFRotation}d respectively. 
	Unlike figures \ref{fig:polarExample} and \ref{fig:polarArray}, the $\Delta \phi_{max}=0.1 \pi$  shift has been subtracted from these data. 
	The magnitude of the black phasor is arbitrary and cannot be compared with the magnitudes of the red phasors. 
	}
	\label{fig:efcPolar}
\end{figure}

A second independent observation that corroborates the existence of an EF is the uniformity of the polar histogram where near EF correction is expected. Figures \ref{fig:efcPolar}a and \ref{fig:efcPolar}b show polar histograms for the cases of no RMP (i.e. no EF correction) and an RMP with $B_{RMP}=0.5$ G at $\phi_{RMP} = 0.33 \pi$. The magnitude of the center-of-mass phasor in the no-RMP case is approximately twice as large as the phasor in the $B_{RMP}=0.5$ G case where the EF is expected to be greatly reduced. This observation is consistent with the prediction of an EF of amplitude similar to the applied RMP, and with phase anti-aligned with this RMP phase.

\section{Discussion}
\label{sec:discussion}

Despite the success of this technique based on the model described in section \ref{sec:model}, it is clear that additional physics is necessary to explain the TM dynamics. The system has been treated as consisting of a $m/n=1/-12$ TM, an applied RMP, and a resonant EF only. In reality, many other TMs and resistive wall modes are present in these discharges and can affect the $m/n=1/-12$ TM through viscous and toroidal coupling. Separately, to decouple the effect of eddy currents in the vacuum vessel from the magnetics, the vessel is assumed to be a first-order low-pass filter. This assumption is only strictly correct when $\dot{\omega}/\omega^2 \ll 1$ \cite{Guo} (in other words, when $\omega$ does not change significantly within one rotation period) whereas it has been observed that $\dot{\omega}/\omega^2 \leq 0.3$ (see figure 3 in reference \cite{Frassinetti2010}). 

\subsection{Comparison of experimental and theoretical $C_1/\tau_R$}

From equation \ref{eq:oneWidth}, and with the readily available experimental measurement of $\omega_o$ from the TM rotation, the phase at which the TM width is maximized (referred to as $\Delta \phi_{max}$) provides an empirical measurement of $C_1/\tau_R$. If the TM rotates quickly relative to the frequency $C_1/\tau_R$, $\Delta \phi_{max}$ goes to $\pi/2$, whereas if the TM rotates relatively slowly, $\Delta \phi_{max}$ goes to zero. Comparing this measurement with a calculated value from our model provides a check on the validity of the model. Also, predictive capability of $\Delta \phi_{max}$ would remove the need to measure it empirically. A poor prediction of $\Delta \phi_{max}$ could cause errors in the passive EF phase identification of up to $\pm \pi/2$; the effects on active EF amplitude and phase identification are not clear.

The value of $\Delta \phi_{max} \approx 0.1 \pi$ measured in section \ref{sec:EF} implies that $ \left (\frac{\omega_o}{C_1/\tau_R} \right ) \approx \tan(0.1 \pi) $. From this and from experimental rotation frequencies $\omega = (3.1-5.7)\times 10^5$ s$^{-1}$, we arrive at an empirical measurement of $C_1/\tau_R= (0.95 - 1.8)\times 10^6$ s$^{-1}$. 

Next we seek an analytic estimate of $C_1/\tau_R$. Using equation \ref{eq:zeroWidth}, we find the following expression for $C_1/\tau_R$,

\begin{equation}
	\frac{C_1}{\tau_R} = \frac{C_0 \eta}{\mu_0 W_0 r_s}
	\label{eq:C1onTauR}
\end{equation}

The ranges we estimate for the parameters on the right hand side of equation \ref{eq:C1onTauR} are shown in the "T2R Range" column of table \ref{tab:tab1}. Some of these parameters have very large experimental uncertainties, particularly $C_0$, $W_0$, and $r_s$. The largest value of the resistivity $\eta$ is derived from Spitzer resistivity with $T_e=200$ eV, $Z_{eff} = 4$, and using the particle trapping correction (see appendix A in reference \cite{LahayeNF15}). Choosing parameters to produce the largest estimate of $C_1/\tau_R$ consistent with geometric constraints and measured perturbed fields (see column T2R in table \ref{tab:tab1}), we find $C_1/\tau_R = 1.1 \times 10^6$ s$^{-1}$. This value is in the experimental range $C_1/\tau_R= (0.95 - 1.8)\times 10^6$ s$^{-1}$ given by the measured $\omega_o$ and $\Delta \phi_{max}$.

\begin{table}
	\centering
    \begin{tabular}{l | c |  c | r }
    	Param. & Unit & T2R Range & T2R  \\
    	\hline
    	$C_0$ & & $20 \pm 20$ & 40   \\
	$W_0$ & cm & $1.5 \pm 1.5 $& 1 \\
    	$T_e$ & eV & $300 \pm 100$ & 200 \\
    	$Z_{eff}$ & & $3 \pm 2$ & 4  \\	
    	$\eta$ & $\mu \Omega \cdot$ m & $1.7 \pm 1.6$ & 3.3  \\
    	$r_s$ & cm & $4 \pm 3$ & 1  \\
    \end{tabular}
    \caption{
    Parameter ranges for EXTRAP T2R and values used for calculations on "T2R". 
    }
    \label{tab:tab1}
\end{table}

\subsection{Applicability to ITER}

The theory used here for classical TMs is not directly applicable to Neoclassical TMs (NTMs) in ITER where $\Delta'(W=0)$ is expected to be negative, and where the bootstrap drive for NTMs cannot be ignored.  However, the EF is still expected to drive some width modulations of a rotating TM in ITER (the islands of interest in ITER might be the  $m/n=$3/2, 2/1, or 3/1). These modulations are expected to be small; for the 2/1 island, $\tau_R \approx 679$ s and $\omega \approx 2\pi \cdot 420$ rad/s, giving $\omega \tau_R \approx 1.8 \times 10^6$ \cite{LaHaye2009}. Modeling similar to that of section \ref{sec:model} could predict the expected modulation of the NTM width,  and help to determine if it will be measurable.

Cylindrical theory suggests that the TM rotation modulates in response to EFs regardless of the value of $\omega \tau_R$ \cite{Fitzpatrick1993}, and therefore the modulated TM rotation might be used for EF identification in ITER. Even when considering the ideal plasma response to a static EF (i.e. the EF goes to zero at the rational surface), the rotation of the TM still modulates due to a torque applied by EF induced currents at the rational surface (see section 4.2 of reference \cite{Fitzpatrick1993}). Modeling should be done to verify that the electromagnetic torque due to an estimated EF is sufficient to produce measurable modulation with the expected viscosity and inertia in ITER. 

\subsection{Future work}

Future EXTRAP T2R experiments might further reduce the errors in figure \ref{fig:intrinsicEF} by using a "binary search" technique. By this we mean using the bifurcation in the span of the resultant phase $\phi_{res}$ (see equation \ref{eq:phiRes}) when the phase of the RMP is scanned at an amplitude above and below the EF amplitude (see figure \ref{fig:0p5G_RMP}) to more precisely determine the EF amplitude. 

Fully passive EF identification should in principle be achievable using the perturbed TM amplitude or perturbed TM velocity data. Passive phase identification of the EF is demonstrated here using the TM amplitude data, assuming prior knowledge of $\Delta \phi_{max}$. The magnitude of the oscillation in the TM amplitude might also be used to identify the EF amplitude passively, though a first attempt using the data in this work was not successful. 

Additional or alternative information could be gained from Fourier-analyzing the magnetic signals: if in the absence of an EF or RMP the TM rotates uniformly at frequency $f$, introducing an EF will make the rotation non-uniform and introduce harmonics of $f$ in the Fourier spectrum. The amplitudes and phases of these harmonics can be used to passively identify the EF. This was not attempted here because while rotation is non-uniform within a rotation period, as required for this technique, it is not sufficiently reproducible from period to period, for Fourier analysis to be applied. An analogous phase shift to $\Delta \phi_{max}$ dependent on the relative sizes of the inertial and viscous torques must be characterized before using the perturbed velocity data passively in this way.

\section{Summary and conclusions}

The amplitude of naturally rotating TMs  is observed to modulate at the TM rotation frequency when a static resonant EF exists in EXTRAP T2R, and the toroidal phase where the TM amplitude is maximized depends on the toroidal phase of the EF (or EF+RMP, if an RMP is applied) \cite{Frassinetti2010}. 

The present work describes a new EF detection technique based on this amplitude modulation.
The technique is developed, validated, and used to identify the $m/n=1/-12$ intrinsic EF in EXTRAP T2R. A simple first-order model is derived from a Modified Rutherford Equation including classical and EF effects, and used to motivate the technique. 

For validation, an RMP of approximate amplitude $10^{-3}$ relative to the equilibrium field was applied and varied in phase. The applied phases were successfully measured up to a constant offset, referred to as $\Delta \phi_{max}$, and an approximately sinusoidal deviation. The constant offset $\Delta \phi_{max}$ was easily characterized by completing a toroidal scan of a moderate amplitude RMP ($\sim2$ to $3\times$ larger than the intrinsic EF) and averaging the measured phases. The complex exponential in equation \ref{eq:oneWidth} is responsible for this offset $\Delta \phi_{max}$ and shows that its value is related to the TM rotation frequency, to the nonlinear correction to the classical stability index, $C_1$, and to a resistive diffusion timescale $\tau_R$. If the TM rotates slowly, $\Delta \phi_{max}$ goes to zero, whereas if the TM rotates quickly, $\Delta \phi_{max}$ goes to $\pi/2$; slow and fast here are relative to the frequency $C_1/\tau_R$. The measured $\Delta \phi_{max}$ suggests a relatively slow TM and is consistent with the theoretical calculation, though the experimental uncertainties in $C_1$ and the resistive diffusion time are large. 

After accounting for $\Delta \phi_{max}$, the approximate sinusoidal deviations are then used to estimate the EF amplitude and phase (figure \ref{fig:2GCentroid}b). An EF of given amplitude and phase produces a unique deviation about $\Delta \phi_{max}$, as shown by equation \ref{eq:phiRes}. Three toroidal scans of an RMP, each with a constant RMP amplitude in the range $B_{RMP}=0$ to 1 G, constrain the EF amplitude to $B_{EF} = 0.7 \pm 0.2$ G and phase to $\phi_{EF} = (1.28\pm 0.08) \pi$ (figure \ref{fig:intrinsicEF}b). This EF estimate is consistent with the highest median TM rotation frequency (figure \ref{fig:EFRotation}d), and the most uniform amplitude behavior (figure \ref{fig:efcPolar}b) when an approximately equal and opposite RMP is applied. 

In summary, in the presence of a naturally rotating tearing mode, this technique can be used in two ways: (1) to passively (i.e. no RMP required) identify the phase of an EF (assuming $\Delta \phi_{max}$ has been characterized), or (2) to detect both the amplitude and phase of an EF by scanning a known RMP in amplitude and phase. The resolution limit of this technique has not been investigated here, but will be the focus of future work.

\section*{Acknowledgements}

R. Sweeney would like to thank W. Choi for useful discussions during the development of this manuscript. This work was partly supported by US DOE Grant DE-SC0008520.

\section*{Appendix A - Coordinate transformations}

Two coordinate systems are used in this work which differ only in their origin. The magnetic sensor positions are defined relative to the machine coordinate system with toroidal coordinate $\phi$. The feedback algorithm with which the RMPs are applied uses a coordinate system relative to the active coil positions with toroidal coordinate $\Phi$. In real space their origins are separated by $5.265^\circ$, but keeping with our toroidal angle convention (section \ref{sec:notations}) we have the relationship $\Phi= \phi - 5.265^\circ \times12= \phi - 67.5^\circ$. All reported RMP phases in this work are mapped to the $\phi$ coordinate system in order to compare with magnetic sensor data. 

\section*{Appendix B - Angle definitions}

Due to the large number of toroidal angle definitions in this work, the following table is provided to consolidate and provide brief descriptions. Please see text for detailed definitions. 

\begin{table}[h]
	\centering
    \begin{tabular}{l |   l }
    	\textbf{Param.} &  \textbf{Definition}  \\
    	\hline
	$\phi^{1,-12}$ & Position of full TM structure \\
	$\phi$ & $12 \times \phi^{1,-12}$, used for model TM (Sec. 3) \\	
	 &  \& unfiltered TM position (Sec. 4.1) \\
	$\underline{\phi}$ & $\phi$ after filtering \\
	$\Delta \phi$ & $\underline{\phi} - \phi_{EF+RMP}$ \\	
	\hline
	$\underline{\phi}^i_{max}$ & $\underline{\phi}$ when amp. is max. in rotation period $i$ \\	
	\hline
	$\phi_{RMP}$ & Position of applied RMP \\
	$\phi_{EF+RMP}$ & Position of EF + RMP \\
	$\phi_{meas}$ & Phase derived from single polar plot \\
	$\phi_{res}$ & $\phi_{meas} - \Delta \phi_{max}$ \\
	\hline
	$\Delta \phi_{max}$ & Average $\Delta \phi$ when TM amp. is max. \\
	\hline
	$\phi_{EF}$ & Position of intrinsic EF \\		
    \end{tabular}
    \caption{
    Brief definitions of all toroidal angle parameters. Angles are grouped by the horizontal lines; first group is fully time-resolved, 
    second group ($\underline{\phi}^i_{max}$ only) is defined on a rotation period, third group is characteristic of a single discharge,
    the fourth group ($\Delta \phi_{max}$ only) is characteristic of a constant-amplitude RMP phase-scan (multiple discharges), and the fifth
    group ($\phi_{EF}$ only) is assumed constant throughout all experiments. All angles except $\phi^{1,-12}$
    span 360$^\circ/12$ in real space. 
    }
    \label{tab:tab2}
\end{table}

\section*{References}
\label{sec:endnotes}

\bibliography{EXTRAP_v2}{}
\bibliographystyle{unsrt}

\end{document}